\documentclass[conference]{IEEEtran}
\IEEEoverridecommandlockouts
\usepackage{amsmath,amsfonts}
\usepackage{algorithmic}
\usepackage{algorithm}
\usepackage{array}
\usepackage[caption=false,font=normalsize,labelfont=sf,textfont=sf]{subfig}
\usepackage{textcomp}
\usepackage{stfloats}
\usepackage{url}
\usepackage{verbatim}
\usepackage{graphicx}
\usepackage{cite}
\usepackage{soul,color}
\usepackage{svg}
\usepackage{tabularray}
\usepackage{caption}
\UseTblrLibrary{booktabs}
\usepackage{wrapfig}
\newcommand{\DM}[1]{\textcolor{black}{#1}}
\newcommand{\LH}[1]{\textcolor{black}{#1}}
\newcommand{\TW}[1]{\textcolor{black}{#1}}
\newcommand{\DMRev}[1]{\textcolor{black}{#1}}
\begin{document}

\title{Constellation Shaping under Phase Noise Impairment for Sub-THz Communications
\thanks{Part of the work done when Dileepa Marasinghe was with Nokia Bell Labs, Stuttgart, Germany. This work has been funded by the European Union’s Horizon Europe research and innovation programme under Hexa-X-II project (grant agreement No. 101095759) and partly by the Research Council of Finland (former Academy of Finland) 6G Flagship Programme (Grant Number: 346208).}}

\author{
\IEEEauthorblockN{Dileepa~Marasinghe$^{1,2}$,
        Le Hang~Nguyen$^{2}$, Jafar~Mohammadi$^{2}$, \\ Yejian~Chen$^{2}$, Thorsten~Wild$^{2}$ and~Nandana Rajatheva$^{1}$}
\IEEEauthorblockA{$^{1}$Centre for Wireless Communications, University of Oulu, Oulu, Finland}
\IEEEauthorblockA{$^{2}$Nokia Bell Labs, Stuttgart, Germany }

\{dileepa.marasinghe, nandana.rajatheva\}@oulu.fi,\\
\{le\_hang.nguyen, jafar.mohammadi, yejian.chen, thorsten.wild\}@nokia-bell-labs.com
}




\maketitle

\begin{abstract}

The large untapped spectrum in the sub-THz allows for ultra-high throughput communication to realize many seemingly impossible applications in 6G. One of the challenges in radio communications in sub-THz is the hardware impairments. 
Specifically, phase noise is one key hardware impairment, which is accentuated as we increase the frequency and bandwidth. 
Furthermore, the \LH{\DMRev{moderate} output power} of the sub-THz power amplifier demands limits on peak to average power ratio (PAPR) signal design. 
\DMRev{Single carrier frequency domain equalization (SC-FDE) has} been identified as a suitable candidate for sub-THz, although some challenges such as phase noise and PAPR still remain to be tackled. In this work, we design a phase noise robust, \DMRev{modest} PAPR 
\DMRev{SC} waveform by geometrically shaping the constellation under practical conditions. We formulate the waveform optimization problem in its augmented Lagrangian form and use a back-propagation-inspired technique to obtain a constellation design that is numerically robust to phase noise, while maintaining a 
 \DMRev{relatively low PAPR compared to the conventional waveforms.}  

\end{abstract}


\section{Introduction}
Realizing sub-THz links envisioned in 6G will favor towards single-carrier (SC) transmissions\cite{HexaXSubTHz}, due to several benefits in terms of PAPR, high coverage, less sensitivity to hardware impairments over the current new radio (NR) multi-carrier (MC) waveforms. The aforementioned characteristics of SC transmissions facilitate enhanced coverage, reliability and most importantly more energy efficient communications, which is one of the essential requirements in the 6G communication networks. 
Among the SC waveform candidates, pure \DM{single carrier frequency domain equalization (SC-FDE)} stands out as a potential 6G waveform  for sub-THz due to the simplest transmitter structure and the \TW{achievable compatibility} with the incumbent NR waveforms \cite{LeHangWaveform}. 

A key hardware impairment that significantly hinders the links in sub-THz is the increasing phase noise (PN), which arises due to the imperfect local oscillators. \LH{PN characteristics are defined by a power spectral density (PSD), which describes the noise power \TW{per Hz} at a certain offset from its operating frequency.}
 \TW{The} \LH{sub-THz frequency is generated by up-converting from the reference frequency where the PN power increases by 6 dB for every doubling of the carrier frequency.}
 \LH{PN can be modelled as a superposition of 2 processes: a correlated Wiener process and an uncorrelated Gaussian process. The contribution of the low frequency Wiener PN is typically addressed by sending phase tracking reference signals (PTRS).} 
 The use of the wide bandwidth for ultra-high data rates\LH{, however, imposes \TW{a} challenge in PN compensation \TW{algorithms} since the contribution of the uncorrelated, and therefore intractable Gaussian PN becomes dominant.}

\begin{figure}[ht]
    \centering
    \includegraphics[width=0.8\columnwidth]{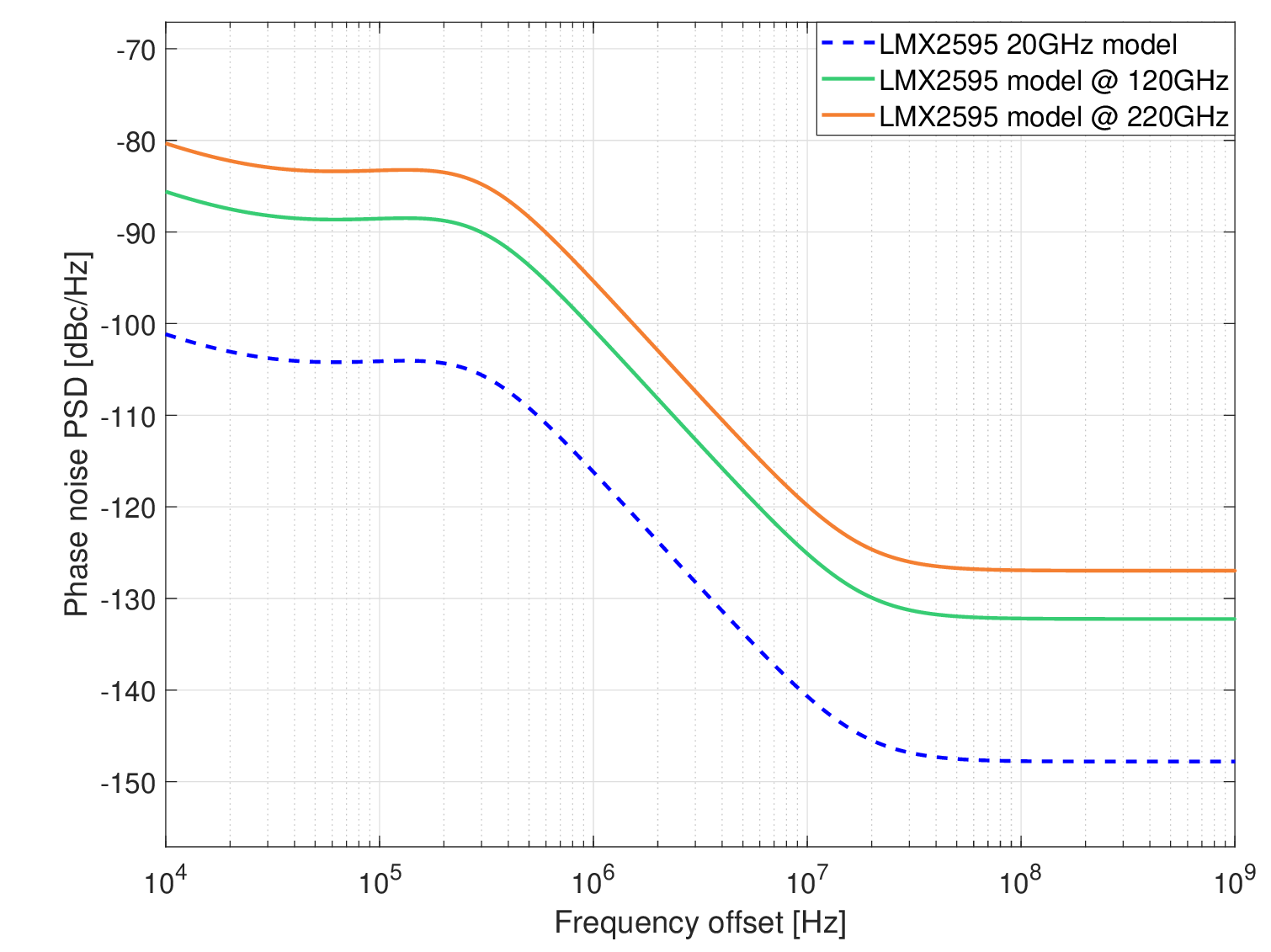}
    \caption{ PSD of the TI LMX2595 oscillator upscaled from 20GHz to 120GHz and 220GHz.}
    \label{fig:phase_noise_compare}
\end{figure}

 An upscaled PSD of the Texas Instrument (TI) LMX2595 oscillator at 120GHz and 220GHz is shown in Figure \ref{fig:phase_noise_compare} which demonstrates the increasing noise power when synthesizing higher frequencies at sub-THz range. Furthermore, note that when a wider bandwidth is employed, the contribution of the noise floor increases, which appear as the Gaussian PN.
 
 \LH{Depending on the constellation pattern used for the modulation, the transmit signals exhibit different robustness against PN, in particularly against the Gaussian PN contribution.} Therefore, a number of works have attempted to design constellations, which are PN-robust \TW{in conjunction with appropriate} detection criteria. Due to limited space, we refer \cite{Foschini,KrishnanPNOptimization,Spiral, CEALetiDigitalDesign},  which are noteworthy among a multitude of other works. An algorithm to select constellation points on a lattice, which minimizes the symbol error probability (SEP) under a Tickonov PN characterization has been presented in \cite{Foschini}. Most of the prior works characterize the PN only with the Gaussian process assuming an ideal estimator, which removes the correlated part of the PN. \DM{Authors in \cite{KrishnanPNOptimization} derived different detection criteria under such Gaussian PN and gradient based optimization was used to optimize the constellation targeting maximum mutual information and symbol error probability. However, the optimization criteria does not account for PAPR. Spiral constellations presented in \cite{Spiral} provided a closed-form method for  defining the constellation, while employing a detection rule originally proposed in \cite{KrishnanPNOptimization}. A more recent work aiming at sub-THz has been presented in \cite{CEALetiDigitalDesign}, which proposed polar-QAM and a method of soft-decision decoding compatible with channel coding based on a detection rule proposed in \cite{KrishnanPNOptimization}}. However, as noted in \cite{CEALetiDigitalDesign}, both spiral constellations and the polar-QAM have similar PAPR characteristics and their performance improvements under PN come at a significant deterioration of the PAPR. \DM{In contrast to these works, we model a practical PN condition, which includes the Wiener part in this work. 
 To the best knowledge of the authors, no work has studied the constellation optimization with PN impairment under a PAPR constraint for sub-THz communications, which is of great interest.}

Recently, constellation shaping has re-gained considerable attention thanks to the available computational capability in developing and deploying artificial intelligence and machine learning (AI/ML) methods \cite{noncovex}. \DM{Significant improvements compared to the conventional transceivers have been demonstrated with learnt waveforms, which include optimized constellations, pulse shaping filters and neural receivers, although they involve a high complexity in the implementation \cite{WaveformFaycal}.} Motivated by these factors, we revisit the constellation design problem for sub-THz in a more practical point of view, constraining the PAPR under practical PN modelling using an SC-FDE waveform \LH{transmission}. We demonstrate through data-driven techniques that it is possible to shape the constellation to be PN-robust, while maintaining a lower PAPR compared to the conventional schemes and deliver better block error rate (BLER) unlocking more potential of an SC waveform for sub-THz communication.

The rest of the paper is organized as follows. We describe our end-to-end system model in Section \ref{3_SystemModel} for the SC-FDE transceiver. Section \ref{4_Problemformulation} explains the problem formulation, while our solution approach is described in Section \ref{5_OptimizationMethodology}. Then we present the results of our evaluation in Section \ref{7_Results} followed by the conclusion and the outlook in Section \ref{8_ConculsiosnsAndOutlook}.


\section{System model}
\label{3_SystemModel}

\begin{figure*}[ht]
    \centering
    \includegraphics[width=0.95\textwidth]{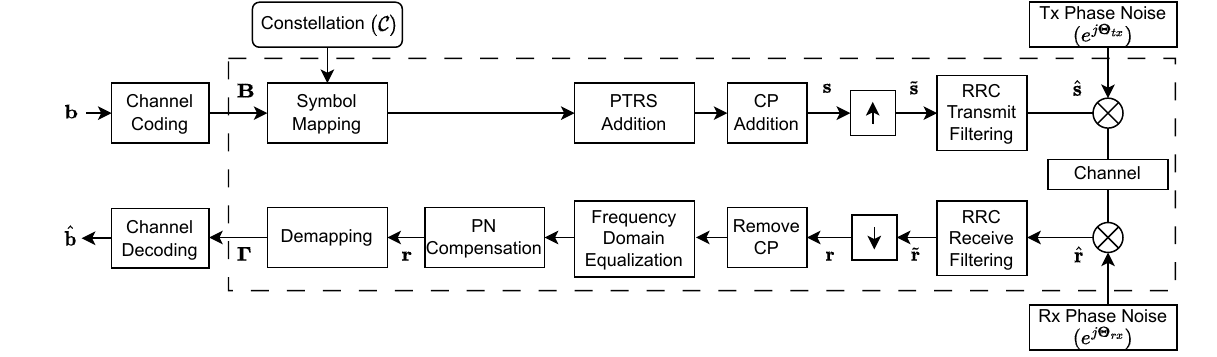}
    \caption{End-to-end system model for the SC-FDE transceiver} 
    \label{fig:ened_to_end_model}
\end{figure*}
In this section, we introduce the discrete-time system model considered in this work. We consider a conventional SC-FDE transceiver as shown in Figure \ref{fig:ened_to_end_model}, which is impaired by phase noise in the transmit and receive sides.  
Let $\mathbf{b}$ be the vector of data bits 
that are to be transmitted through the system. First, $\mathbf{b}$ is channel coded with rate $r<1$ \DM{to a} coded matrix of bits ${\mathbf{B}}_{N_D \times K}$ where $K$ is the \DM{number of bits per symbol} and $N_D$ is the number of \LH{modulated} data symbols. 
Then, each column of $K$ bits are mapped to the corresponding symbol based on a constellation $\mathcal{C} \in \mathbb{C}^{2^K} $ with the respective bit-labelling following the \DMRev{well-known} bit-interleaved coded modulation (BICM) scheme. 
\DM{Next, PTRS time samples are organized into $Q$ groups and evenly distributed between the mapped data symbols, with each group containing $N_P$ time samples.}
\DM{They are primarily intended for compensating the Wiener part of the PN at the receiver}. This is followed by an $N_{CP}$-cyclic prefix (CP) extension to the block to form the transmit symbol vector $\mathbf{s} \in \mathbb{C}^{N}$, where $N = N_D + Q \times N_{P} + N_{CP}$, is the length of the transmission block. Then $\mathbf{s}(n)$ is upsampled with an oversampling factor of $M$ compared to the symbol rate such that, 
\begin{equation}
\Tilde{\mathbf{s}}(n)=\begin{cases} \begin{aligned} 
    &\mathbf{s}\left(\frac{n}{M}\right), \quad \frac{n}{M} \in \mathbb{Z},\\
    & 0, \quad\quad\quad\quad \text{otherwise.}
\end{aligned} \end{cases}
\end{equation}
The upsampled signal \LH{$\Tilde{\mathbf{s}}(n)$} is filtered by a root-raised-cosine (RRC) transmit filter $\mathbf{g}_{RRC}$ with a roll-off factor $\beta$ as,
\begin{equation}
    \hat{\mathbf{s}}(n) = \Tilde{\mathbf{s}}(n) \ast \mathbf{g}_{RRC}(n).
\end{equation} \DM{where $\ast$ denotes the convolution.} The filtered signal $\mathbf{\hat{s}}(n)$ is then sent to the analog front end (AFE) of the transmitter, where the \TW{phase jitter} of the local oscillator impairs the signal with PN samples $\mathbf{\Theta}_{tx}(n)$. Then the PN impaired transmit signal goes though the channel. 
\LH{Since the focus is on the PN impairment, we emulate a scenario of perfect channel equalization by deploying an AWGN channel.}
The received signal at the AFE of the receiver is further impaired by the receiver phase noise samples $\mathbf{\Theta}_{rx}(n)$. The sampled received signal $ \hat{\mathbf{r}}(n)$
can be written as, 
\begin{equation}
     \hat{\mathbf{r}}(n) = \hat{\mathbf{s}}(n) e^{j\mathbf{\Theta}_{tx}(n)} e^{j\mathbf{\Theta}_{rx}(n)} + \mathbf{w}(n)
\end{equation}
where $\mathbf{w}(n) \sim \mathcal{CN}(0,\,\sigma^{2}\mathbf{I})$, denotes the \DMRev{complex} AWGN with variance $\sigma^2$. 
The sampled signal is then filtered with an RRC receive filter with the same roll-off $\beta$ as,
\begin{equation}
    \Tilde{\mathbf{r}}(n) = \hat{\mathbf{r}}(n) \ast \mathbf{g}_{RRC}(n)
\end{equation}
and downsampled by a factor of $M$, which provides the received symbols $\mathbf{r}(n) = \Tilde{\mathbf{r}}(nM)$.

First, the CP is removed from $\mathbf{r}(n)$.  The frequency domain equalization step is transparent thanks to the AWGN channel deployment. \LH{PTRS time samples are then extracted for PN compensation.}
We describe this PN compensation in more detail in subsection \ref{phase_noise_compensation}.
The resulting data symbol block ${\mathbf{r}}(n)$  is sent to the demapper to generate the log-likelihood ratios (LLRs) for the bits transmitted  with the conventional AWGN optimal demapper where the demapper treats the signals to be only corrupted by AWGN noise. \DMRev{Therefore, this is a mismatched demapper under the additional residual phase noise. Our aim is to design the constellation such that, it is compatible with this demapper.} We consider bit-metric decoding (BMD) where LLR for each bit is  calculated\DM{\footnote{Tensorflow single-precision implementation of LogSumExp \texttt{tf.math.reduce\_logsumexp} was used for computation.}} as,
\begin{equation}
    \Gamma_{n,k} = \log \left( \frac{\sum_{c_i\in \mathcal{C}(1)} \exp \left(-\frac{|{\mathbf{r}}(n)-c_i|^2}{\sigma^2}\right)}{\sum_{c_i\in \mathcal{C}(0)} \exp \left(-\frac{|{\mathbf{r}}(n)-c_i|^2}{\sigma^2}\right)} \right)
\label{eq_AWGN_demapper}
\end{equation}
where $\Gamma_{n,k}$ is the LLR for the $k^\text{th}$ bit $(0 \leq k \leq K - 1)$ of the $n^\text{th}$ symbol $(0 \leq n \leq N_D - 1)$, and $\mathcal{C}(0)$ and $\mathcal{C}(1)$ are the subsets of $\mathcal{C}$, which contains all constellation points with the $k^\text{th}$ bit label set to $0$ and $1$ respectively. Note that the definition of the LLR here is the inverse of the conventional definition, which is intentional to align with the logit definition commonly used in the machine learning frameworks.

\subsection{Phase noise model}
The main focus in this work is to design a symbol constellation, which is robust \LH{against PN.} Hence, accurate modelling of the phase noise is of paramount importance. \DM{Furthermore, our solution is data-driven, which allows us to approximate the \LH{practical} PSD \LH{characteristics} of the PN to a sufficient accuracy in contrast to models that neglect the Wiener part based on an ideal estimator assumption \cite{KrishnanPNOptimization},\cite{CEALetiDigitalDesign}.} Therefore, we consider the \LH{well-established} multi pole-zero \LH{PN} model \LH{ recommended by} \DMRev{\cite[\S6.1.10]{3gpp.38.803}}. The PSD we consider is given by,
\begin{equation}
S\left( f \right) = \text{PSD0}\frac{\prod\limits_{n = 1}^N 1 + {\left( {\frac{{{f}}}{{{f_{z,n}}}}} \right)^{{\alpha _{z,n}}}}} {\prod_{n=1}^{N} 1+\left(\frac{f}{f_{p, n}}\right) ^{\alpha_{p, n}}}
\end{equation}
where \DM{$\text{PSD0}$} is the power spectrum at zero frequency, $f_{z,n}$ and $f_{p,n}$ are zero and pole frequencies. $\alpha_{z,n}$ and $\alpha_{p,n}$ are the powers at each zero and pole frequencies. \DMRev{Phase noise samples are generated from the parameters, using the relevant models specified in Table \ref{sim_table}. Then the generated phasors are multiplied with the oversampled pulse-shaped signal which implicitly simulates the inter symbol interference (ISI) caused by the phase noise. }  

\subsection{\LH{Phase noise compensation}}\label{phase_noise_compensation}
\begin{figure}[]
    \centering
    \includegraphics[width=0.5\columnwidth{}]{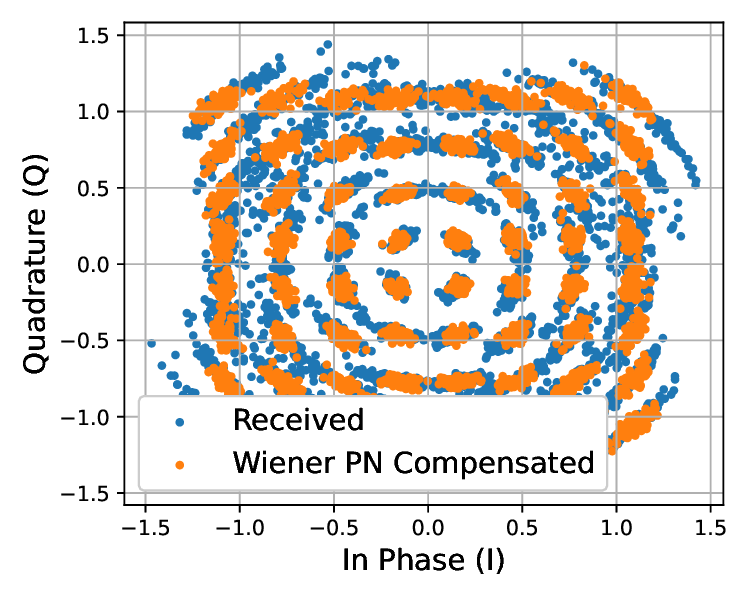}
    \caption{ \DMRev{PN impacted received signal \TW{before} compensation and the signal after Wiener PN compensation, which shows the residual PN. ($E_b/N_0 = 25 \:dB$)}}
    \label{PNCompensated}
\end{figure}

A possible approach for the phase noise compensation is to average the phase error over a group of PTRS samples, then interpolate over the block for cancelling the phase error \cite{PN_tracking_Linear}.
The average phase error estimate over the $q^\text{th}$ group $\Bar{\Theta}_q$, is calculated as,
\begin{equation}
    \DMRev{\Bar{\Theta}_q = \text{arg} \Bigg(\frac{1}{N_P} \sum_{ m = 0}^{N_P-1} \frac{\mathbf{r}_q(m)\mathbf{p}^{\ast}_q(m)}{{|\mathbf{p}_q(m)|}^2}\Bigg)} 
\end{equation}
where $\text{arg}(\cdot)$ is the angle of the complex value, $\mathbf{r}_q$ are the received phase tracking samples in the $q^\text{th}$ group and $\mathbf{p}_q$ are the corresponding \LH{transmitted} PTRS samples. Then, the average values of all $Q$ groups can be interpolated to generate a phase error estimate $\mathbf{\Bar{\Theta}}$ of length $N_D + Q \times N_{P}$. The estimate is used to compensate for the PN by multiplying with the received signal as $\mathbf{r}(n)e^{-j\Bar{\mathbf{\Theta}}(n)}$.
\LH{Note that this method mainly removes the correlated PN contribution, which can readily be tracked. However, a residual amount of the PN remains uncompensated whose quantity depends on the occupied bandwidth and the uncorrelated PN level of the local oscillator.} One could increase the PTRS density to \TW{better} compensate for the PN errors, but since the Gaussian PN is uncorrelated, the pilot density \TW{would need to be} be increased significantly to fully compensate for the PN, which means useful resources for the data transmission are decreased by a considerable amount. Figure \ref{PNCompensated} \DMRev{shows the comparison between the PN impaired received signal and the resulting signal with} residual phase noise \DMRev{after PTRS based Wiener PN compensation} for a 64QAM transmission at 120 GHz with 3.93 GHz bandwidth, which is coming from an LMX2595 oscillator PN model\DMRev{\cite{ti_lmx2595_datasheet}} at the transmitter side and a 3rd generation partnership project (3GPP) user equipment (UE) model 1\DMRev{\cite{3gpp.38.803}} at the receiver side. As mentioned earlier also, one of our main objectives in this work is to ensure robustness of the waveform to these residual phase errors.

\section{Problem formulation}
\label{4_Problemformulation}
In this section, we describe the optimization problem for the constellation shaping using the system model in Section \ref{3_SystemModel}. We consider the SC system in the Figure \ref{fig:ened_to_end_model}, which the constellation is trainable. 
 This essentially means we optimize the conventional SC-FDE system for an optimized constellation to maximize the information rate under practical constraints. It has been shown that \DMRev{maximizing} the \TW{bit-metric decoding} rate \DMRev{can be achieved equivalently by minimizing the} total binary cross entropy (BCE) loss \cite{FaycalPilotless}, which is commonly used in binary classification problems. The total binary cross entropy loss ${\mathcal{ L}}$ is expressed as,
\begin{align}
\begin{split}
\mathcal{ L}({\mathcal{ C}} )= {}& -\frac {1}{N_D} \sum _{n=0}^{N_D-1} \sum _{k=0}^{K-1}  b_{n,k} \mathop {\mathrm {log_{2}}} (\Gamma_{n,k} ) \\ 
&+ (1-b_{n,k}) \mathop {\mathrm {log_{2}}} (1-\Gamma_{n,k} )  \\
\end{split}  
\end{align}
where $b_{n,k}$ and $\Gamma_{n,k}$ are the corresponding bits and the LLRs, 
when the constellation $\mathcal{ C}$ is used.

Due to current technological \LH{challenges} at sub-THz frequencies, the PA's \LH{output power and} non-linearity depicts a larger detrimental impact on the energy efficiency of the radio system, in comparison to the current lower frequencies. 
Thus, the PA must be operated \LH{at the least possible power backoff} 
in order to maintain a meaningful communications range and efficiency. \DMRev{In this work, we assume the operation is in the linear region of the PA, and target a reduced PAPR to lower the backoff.}
Therefore, our optimization problem is to learn the constellation in an end-to-end fashion by minimizing the BCE loss subject to a 
PAPR limit, 
\LH{in} the presence of phase noise. 
Formally this could be stated as,
\begin{subequations}\label{optimization_problem}
\begin{align}
&\underset{\mathcal{C}}{\text{minimize}}~\mathcal{ L}({\mathcal{ C}}  ) \\ 
&\text{subject to}~\mathbb{E}_{c_i \in\mathcal{C}} \left[ c_i \right] = 0 \label{const_mean_condition}\\
&\hphantom{\text{subject to}~}\mathbb{E}_{c_i \in\mathcal{C}} \left[\left| c_i \right|^2\right] = 1 \label{const_energy_condition}\\
&\hphantom{\text{subject to}~} \text{PAPR}(\mathcal{C}) \leq \epsilon_p \label{papr_constraint}
\end{align}
\end{subequations}
The condition in \eqref{const_mean_condition} ensures centred constellation on the IQ plane while \eqref{const_energy_condition} ensures that the learnt constellation is normalized in energy. The PAPR is a function of the constellation points enforced in the condition in \eqref{papr_constraint} to be 
\LH{below} $\epsilon_p$. 
 
\subsection{Trainable constellation}
A trainable constellation, which uses 2D signalling points on the IQ-plane can be modelled by a set of ${2^K}$  trainable complex weights $\tilde {\mathcal{C}}$ where $K$ is the number of bits.
Such a constellation $\mathcal{C}$ is described by \cite{FaycalPilotless,WaveformFaycal},

\begin{equation} 
    \mathcal{C}= \frac{\tilde {\mathcal{C}} - \frac{1}{2^{K}}\sum_{c_i \in \tilde {\mathcal{C}}}^{}c_i}{\sqrt{\frac{1}{2^{K}}|\tilde {\mathcal{C}} - \frac{1}{2^{K}}\sum_{c_i \in \tilde {\mathcal{C}}}^{}c_i|^2}} 
\label{eq_trainable_constellation}
\end{equation}
The bit labelling is set by the binary value of the index of each element in $\mathcal{C}$. The bit to symbol mapping is done \LH{as in} a conventional constellation. Therefore, no additional complexity is introduced when it is used in the transmitter compared to the traditional mapping. However, it allows us to optimize the constellation points and the corresponding bit-labels. During training, the constellation points will traverse on the IQ-plane and rearrange themselves optimally to meet the optimization criteria.

\subsection{Constraints}
We focus on the constraints of the problem in \eqref{optimization_problem}. The conditions \eqref{const_energy_condition}, \eqref{const_mean_condition}  are implicitly enforced by the model for the constellation. The power characteristics of the waveform depends on the distribution of the constellation points as well as the pulse shaping filters. Here we use root-raised cosine (RRC) filters as the transmit and receive filters. Note that phase noise does not change the power of the transmit signal since it distorts only the phase. The PAPR is best described in a probabilistic sense \cite{PAPR} as,
\begin{equation}
    \DMRev{\text {PAPR} = {\text{min}} \bigg\{ \nu \: \big| \:\Pr \left ({\frac {{|\mathbf{\hat{s}}(n)|}^2}{ \mathbb{E}[{|\mathbf{\hat{s}}(n)|}^2]} > \nu }\right) \leq \delta_p \bigg\}}
\end{equation}
The \DMRev{PAPR} in its conventional meaning can be obtained by setting $\delta_{p} = 0$.
In the PAPR constraint in \eqref{papr_constraint}, we aim to have $\text{PAPR}(\mathcal{C}) \leq \epsilon_P$, which means we ensure that the threshold $\nu $ is our desired PAPR $\epsilon_p$, which can be equivalently expressed as, 
\begin{equation}
    \mathbb{E}\bigg( \text{max}\Big(\frac {{|\mathbf{\hat{s}}(n)|}^2}{ \mathbb{E}[{|\mathbf{\hat{s}}(n)|}^2]}-\epsilon_p, 0 \Big)\bigg)  = 0
\end{equation}
where the expectation is over the transmit signal $\mathbf{\hat{s}}(n)$. Therefore the PAPR constraint, $\Psi$ can be computed by Monte Carlo sampling of the transmit signal $\mathbf{\hat{s}}(n)$ during training as,
\begin{align}
    \Psi (\mathcal{C}, \epsilon_p) &\approx \frac{1}{V}\sum_{v=1}^{V} \text{max}\Big(\frac {{|\mathbf{\hat{s}}_v(n)|}^2}{ \mathbb{E}[{|\mathbf{\hat{s}}(n)|}^2]}-\epsilon_p, 0 \Big) \notag\\
    &= 0
\end{align}
where $V$ is the number of power samples.


\section{Optimization Methodology}
\label{5_OptimizationMethodology}
\label{Section_5_Optimization}
\DM{The problem of constellation shaping for maximizing the information rate itself is a non-convex problem \cite{noncovex}. Therefore, the problem we are trying to solve is a non-convex constrained optimization problem.} Since the goal is to optimize the system in an end-to-end fashion, we employ the augmented Lagrangian method for solving the problem, which allows reformulating the BCE loss and the constraints in the problem \eqref{optimization_problem} into a differentiable loss function. With the construction of the augmented Lagrangian, the constrained optimization problem can be solved with a series of unconstrained problems using stochastic gradient descent (SGD) by computing the gradients and back propagating through the system with respect to the trainable parameters. This optimization method has been successfully used in \cite{WaveformFaycal},\cite{MathieuConference} for similar problems and known to be converging faster compared to the penalty method. The augmented Lagrangian can be written for $\lambda > 0$ as,
\begin{equation}    
\label{augmented_lagrangian_reform}
\mathcal{ L}_{aug}({\mathcal{ C}}, \mu_P, \lambda ) =  \mathcal{ L}({\mathcal{ C}} ) 
+ \mu_P\Psi (\mathcal{C}, \epsilon_p)  + \frac{\lambda}{2} |\Psi (\mathcal{C}, \epsilon_p)|^2
\end{equation}
where $\mu_P$ is the Lagrangian multiplier for the PAPR constraint and $\lambda$ is the penalty parameter. 

\subsection{System training}
Since we apply tools from AI/ML community, we use the term ``training'' and optimizing throughout this paper interchangeably, although, we do not train any neural network.
The optimization procedure involves iterative minimization of the augmented Lagrangian in \eqref{augmented_lagrangian_reform} by SGD, which we outline in \LH{ the following Algorithm  \ref{train_alg}}.
The initial $\textit{System Parameters}$ include the carrier frequency and bandwidth, number of bits $K$, data block length $N_D$, PTRS parameters $Q$ and $N_P$, CP length $N_{CP}$, oversampling factor $M$, filter lengths $L$, and the phase noise model parameters on Tx and Rx sides.  \LH{The end-to-end model comprises all the functionality inside the dashed box indicated in Figure \ref{fig:ened_to_end_model}}. During the SGD steps, randomly generated data bits are carried through the end-to-end model to generate bitwise LLRs, which are used to calculate the BCE loss. Then we have a fully differentiable loss function which is used to compute gradients w.r.t. each of the trainable parameters and back propagated through the system to update the trainable parameters.
\begin{algorithm}[ht]
\caption{Training algorithm}\label{train_alg}
\begin{algorithmic}
\STATE $\textbf{Set} \;\; \textit{\{System Parameters\}}$
\STATE $\textbf{Initialize} \;\; {\mathcal{ C}}, \mu_P^{[0]},\lambda^{[0]} $ 
\STATE $\textbf{for} \;\; i = 0, \dots \textbf{do}$
\STATE \hspace{0.5cm}$ \textbf{Perform SGD on} \: \mathcal{ L}_{aug}({\mathcal{ C}}, \mu_P^{[i]}, \lambda^{[i]} )$ 

\STATE \hspace{0.5cm}$\triangleright \: \text{Update Lagrangian multiplier}$
\STATE \hspace{0.5cm}$ \textbf{Recompute}: \Psi^{[i]} (\mathcal{C} , \epsilon_p) $
\STATE \hspace{0.5cm}$\mu_P^{[i+1]}\leftarrow \mu_P^{[i]} + \lambda^{[i]} \Psi^{[i]} (\mathcal{C}, \epsilon_p) $
\STATE \hspace{0.5cm}$\triangleright \: \text{Update penalty parameter}$
\STATE \hspace{0.5cm}$\lambda^{[i+1]} \leftarrow \tau\lambda^{[i]} \: \text{where} \:\tau>1$
\STATE $\textbf{end for}$
\end{algorithmic}
\label{alg1}
\end{algorithm}


\section{Evaluation and Results}
\label{7_Results}
\begin{figure*}[ht]
    \centering
    \includegraphics[width=\textwidth]{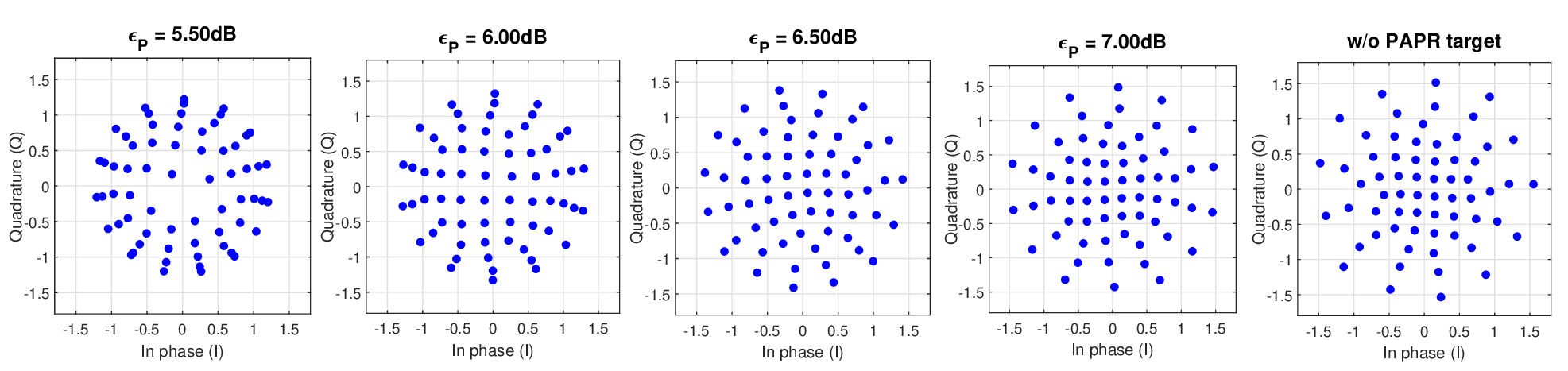}
    \caption{Learned constellations under different PAPR targets and without a PAPR target}
    \label{fig:constellations}
\end{figure*}
\subsection{Evaluation setup}
The implementation was done in the widely-used machine learning framework Tensorflow and some components from the open-source library for physical layer research – Sionna \cite{hoydis2023sionna}. We use the automatic differentiation feature and the optimizers to implement the SGD step. In Table \ref{sim_table}, we summarize the simulation parameters used for the evaluation. The constellation is initialized with a standard-QAM constellation with Gray-labelling.  We train the end-to-end system on a range of $\frac{E_b}{N_0}$ values with uniform sampling at each iteration such that the learned waveform is robust in the  applicable SNR range for a chosen modulation order. \DMRev{In terms of deployment, note that our intention is to learn the constellations offline, and deploy in the real scenario rather than online adaptation.} The AWGN noise variance $\sigma^2$ for a certain $\frac{E_b}{N_0}$ is calculated as,
\begin{equation}
    \sigma^2 = \Bigg( \Big(\frac{E_b}{N_0}\Big) r M \Big(\frac{N-QN_P}{N+N_{CP}}\Big)\Bigg)^{-1}
\end{equation}
where $r$ is the code rate. 
The end-to-end system is trained as an uncoded ($r=1$) system and for evaluation, an outer channel code with 5G NR LDPC was used for computing the BLER of the system. 



\begin{table}[ht] \caption{Simulation parameters}\label{sim_table}
\begin{tabular}{|l|l|}
\hline
\textbf{Carrier frequency}                                                        & 120 GHz                                                                    \\ \hline
\textbf{Bandwidth}                                                                 & 3.93 GHz                                                                   \\ \hline
\textbf{Block size$(N)$}                                                           & 4096                                                                          \\ \hline
\textbf{Bits per symbol $(K)$}                                                    & 6                                                                                  \\ \hline
\textbf{Oversampling factor  $(M)$}                                                 & 4                                                                                     \\ \hline
\textbf{RRC filter span  $(L)$}                                                     & 32 symbols                                                                            \\ \hline
\textbf{RRC roll-off   $(\beta)$}                                                     & 0.3                                                                           \\ \hline
\textbf{CP ratio   $(N_{CP}/(N+N_{CP}))$}                                           & 7.03125 \%                                                                \\ \hline
\textbf{Tx Phase noise model}                                                       & TI LMX2595 @20GHz (upscaled) \DMRev{\cite{ti_lmx2595_datasheet}}                                                         \\ \hline
\textbf{Rx Phase noise model}                                                       & \begin{tabular}[c]{@{}l@{}}3GPP TR38.803 v14.2.0 \\ UE model 1 \DMRev{\cite[Table 6.1.11.2-1]{3gpp.38.803}}\end{tabular} \\ \hline
\textbf{No. of PTRS groups  $(Q)$}                                                          & 32                                                                                 \\ \hline
\textbf{\begin{tabular}[c]{@{}l@{}}PTRS samples per\\ PTRS group $(N_P)$\end{tabular}}& 4                                                                                     \\ \hline
\textbf{PTRS samples}                                                               & Zadoff-Chu sequences                                                                  \\ \hline\hline
\textbf{PAPR targets $(\epsilon_p)$}                                                & 5.5 dB, 6.0 dB, 6.5 dB, 7.0 dB                                         \\   \hline
\DMRev{\textbf{No. of power samples (V)} }                                                                & 4e5                                                                                    \\ \hline
\textbf{Batch size}                                                                 & 10                                                                                    \\ \hline
\textbf{Optimizer}                                                                  & Adam                                                                                  \\ \hline
\textbf{Learning rate}                                                              & 1e-3                                                                                  \\ \hline
\textbf{Training $E_b/N_0$ range}                                                        & {[}8-20{]} dB                                         \\ \hline
\textbf{$E_b/N_0$ sampling}                                                              & Uniform                                                                               \\ \hline \hline
\textbf{Evaluation – Channel code}                                                  & 5G NR LDPC                                                                               \\ \hline
\textbf{Code rate}                                                                  & 3/4                                                                                   \\ \hline
\textbf{BP iterations}                                                              & 50                                                                                    \\ \hline
\end{tabular}
\vspace{-5pt}
\end{table}

\subsection{Results}
\begin{figure*}[ht]
    \centering
    \includegraphics[width=0.9\textwidth]{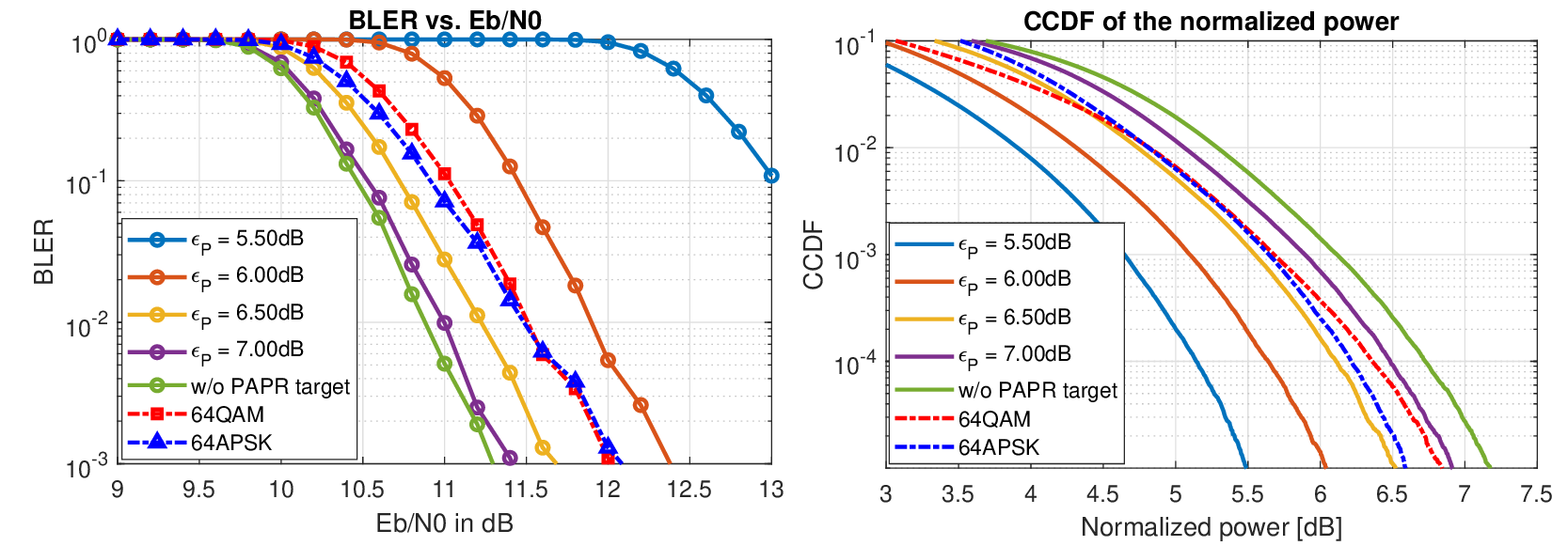}
    \caption{BLER performance and CCDF of normalized power of the learned schemes}
    \label{fig:performance}
\end{figure*}

Figure \ref{fig:constellations} shows the learned constellations under the different PAPR targets \DMRev{and the case without a PAPR restriction}, while their BLER performance and the CCDF of the normalized power distribution are shown in Figure \ref{fig:performance}. We also show the two baseline solutions with the conventional gray labelled QAM and 8+16+20+20 APSK constellation. The latter \LH{ is chosen to be deployed in the digital video broadcasting standard by ETSI (DVB-S2)\cite{standard2020digital}} and was found to be the best performing APSK scheme in terms of both BLER and PAPR for our setting.
When considering the constellations, we can see a distinctive branching effect in all the cases in the outer regions of the constellation. 
Since the constellation is trained for the conventional Euclidean distance based demapping, the points arrange themselves such that they are apart enough in the outer regions, 
which causes the branching effect. The reason is, that the phase noise affects more for the constellation points in the outer part compared to the inner points with the increasing distance from the center (see Figure \ref{fig:constellations}). 
\DM{Additionally, it is noticeable, \TW{while} the PAPR constraint becomes more stringent, the maximum amplitude of the constellation points decreases, and the points pack themselves closer together. However, this comes at the cost of BLER performance as it impacts decoding accuracy.} In the case of $\epsilon_p=5.5 \: \text{dB}$ the constellation attains the ``dough-nut" like shape restricting the amplitude variation to achieve lower PAPR.
Nevertheless, the BLER is significantly high. Relaxing the \DMRev{PAPR} constraint with $\epsilon_p = 7.0 \: \text{dB}$, we can observe more gain in terms of BLER. \DMRev{When the PAPR restriction is removed and trained under BCE only, the learned constellation achieves a PAPR around $7.2\: \text{dB}$ at $10^{-5}$ CCDF level while giving the highest BLER gain.} For the evaluated setup, 
with $\epsilon_p = 6.5 \: \text{dB}$, compared to 64-QAM we achieve 0.3 dB performance improvement at 1\% BLER and a 0.3 dB better PAPR. Compared to the 64-APSK case, the PAPR improvement is 0.1 dB with a similar gain in BLER. Additionally, the learned constellations achieved Gray-labelling, which is not shown for clarity. 





\section{Conclusion}
\label{8_ConculsiosnsAndOutlook}
Constellation shaping under phase noise impairment and PAPR constraints is investigated in this paper in an end-to-end SC-FDE transceiver for sub-THz communications.
The results demonstrate that the utilized data-driven technique allows us to
identify an optimized constellation geometry, which provides performance gain in terms of both PAPR and BLER under practical constraints compared to the conventional approaches.
Moreover, this allows us to have a trade off between the PAPR and BLER performance. The learned constellation can directly be utilized in the SC-FDE transceiver by simply replacing the QAM or APSK constellations without adding any other complexity. Further PAPR reduction and BLER gain are possible by adapting the pulse shaping filter pair and exploring other demapping techniques, which is the main scope of the future work. 




 
 \bibliographystyle{IEEEtran}
\bibliography{bibi}

\vfill

\end{document}